\documentclass{PoS}

\title{Possible resonance effect of dark matter axions in SNS Josephson junctions}

\ShortTitle{Possible resonance effect of dark matter axions in SNS Josephson junctions}

\author{\speaker{Christian Beck}
\\
        School of Mathematical Sciences, Queen Mary University of London, London E1 4NS, UK\\
        E-mail: \email{c.beck@qmul.ac.uk}}

\abstract{Dark matter axions can generate peculiar effects in special types of Josephson junctions, so-called SNS junctions. One can show that the axion field equations in a Josephson environment allow for very small oscillating supercurrents, which manifest themselves as a tiny wiggle in the I-V curve, a so-called Shapiro step, which occurs at a frequency given by the axion mass. The effect is very small but perfectly measurable in modern nanotechnological devices. In this paper I will summarize the theory and then present evidence that candidate Shapiro steps of this type have indeed been seen in several independent condensed matter experiments. Assuming the observed tiny Shapiro steps are due to axion flow then these data
         point to an axion mass of $(106 \pm 6)\mu$eV, consistent with
         what is expected for the QCD axion. In addition to the above small Shapiro resonance effects at frequencies in the GHz region one also expects to see broad-band noise effects at much lower frequencies. Overall this approach provides a novel pathway for the future design of new types of axionic dark matter detectors. The resonant Josephson data summarized in this paper are consistent with a 'vanilla' axion
 with a coupling constant $f_a=\sqrt{v_{EW}m_{Pl}}=5.48 \cdot 10^{10}$GeV given by the geometric average of the electroweak symmetry breaking scale $v_{EW}$ and the Planck mass $m_{Pl}$.}

\FullConference{The European Physical Society Conference on High Energy Physics\\
		5-12 July, 2017\\
		Venice}

\begin{document}

\section{Introduction}\label{sec1}

Besides WIMPs, the QCD axion is one of the leading candidates to account for dark matter in the universe \cite{prl}--\cite{organ}.
In recent years some new experimental techniques have been proposed to detect the axion in new types of laboratory experiments on the Earth
\cite{prl,pdu,c0,c1,c2,e1,e2,10a,fluxnoise}.
One of these detection ideas is based on resonance effects in special types of Josephson junction devices.
In this paper I will summarize the most important results based on the idea that so-called SNS Josephson junctions
could be highly effective axion detectors\cite{prl,pdu}. I will briefly discuss the 'axionic Josephson effect' that underlies this proposal and then come to experimental consequences and observations.
Remarkably, out of this approach comes a concrete prediction of the axion mass, taking into account observed resonance
effects that were seen in experiments done by various condensed matter groups in the recent past (and that were re-interpreted in terms of axion physics in \cite{prl,pdu}):
 If these data are produced by axions then the axion mass must be in the region 100...112 $\mu$eV. This gives
 future axion haloscope experiments (e.g. \cite{rybka,madmax,organ}) a powerful suggestion of where to start the search and where one might ultimately
 be successful in finding the QCD axion.

The prediction that the QCD axion mass could be in this region was published for the first time in 2013 in \cite{prl}, at a time where most people thought the axion mass would be
smaller (in the range of a few $\mu$eV as searched for by the ADMX experiment at that time).
Later, in 2015 Kawasaki
et al. \cite{saika1} obtained a prediction consistent with the above region (their 2015 paper quotes the result $60 \mu  eV<m_ac^2 <150\; \mu eV$).
Their numerical method is based
on a completely different method, namely estimating the production rate of axions from simulations of domain walls
and strings and comparing with the measured abundance of dark matter, assuming all of dark matter is made up of axions.
The recent lattice simulations of Borsanyi et al. \cite{nature} confirmed this rather large value of the axion mass
(they quote 50...1500$\mu$eV), and also theoretical models such as the so-called SMASH model \cite{smash} advocate an axion mass of around $50...200\mu$eV. On the other hand,
axionic string and domain wall simulations are very complicated and a modified numerical method \cite{moore} taking into account large string tensions very recently yielded a somewhat different result than \cite{saika1},
with a tendency
that axion production from strings and walls is somewhat less efficient, thus in principle
allowing for a smaller axion mass, assuming that all of dark matter is axions.


In the following we give a brief summary of the most important theoretical results of \cite{prl,pdu} and then come
to experimental predictions, in particular we give an update on the axion mass prediction,
  taking into account further data from other condensed matter experiments
in addition to the single experiment discussed in \cite{prl}. The prediction we arrive at by taking the average of 5 independent condensed matter experiments \cite{prl,pdu,ex1,ex2,ex3,ex4,ex5} is
\begin{equation}
m_ac^2=(106 \pm 6) \mu eV.
\end{equation}

\section{The axionic Josephson effect}

Let us first recall some basic facts about Josephson junctions which are needed in the following.
A Josephson junction consists of two superconductors separated by a weak-link region.
The weak link-region is an insulator for tunnel junctions and a normal metal for so-called SNS junctions.
The distance between the superconducting plates is $d\sim 1nm$ for tunnel junctions, and about $d\sim 1\mu m$ for SNS
junctions. If a DC voltage $V$ is applied then a Josephson junction emits Josephson radiation of frequency $\hbar \omega_J=2eV$.
In this case the phase difference $\delta (t)$ of the macroscopic wave functions
 describing the 'left' and 'right' superconductor grows linearly in time: $\delta :=\Phi_L-\Phi_R \sim \omega_J t$.
 A Josephson junction in that state was proposed as an axion detector in \cite{prl}, provided the Josephson frequency $\omega_J$
 resonates with the axion mass.

The field equations of axions passing through a Josephson environment are \cite{pdu}
\begin{eqnarray}
\ddot{\theta} + \Gamma \dot{\theta} -c^2 \nabla^2 \theta +\frac{m_a^2c^4}{\hbar^2} \sin \theta
&=& - \frac{g_\gamma}{4 \pi^2} \frac{1}{f_a^2} c^3 e^2 \vec{E} \vec{B} \\
\nabla \times \vec{B} - \frac{1}{c^2} \frac{\partial \vec{E}}{\partial t} &=& \mu_0 \vec{j} +\frac{g_\gamma}{\pi}
\alpha \frac{1}{c} \vec{E} \times \nabla \theta - \frac{g_\gamma}{\pi}\alpha \frac{1}{c} \vec{B} \dot{\theta} \\
\nabla \vec{E} &=& \frac{\rho}{\epsilon_0} + \frac{g_\gamma}{\pi} \alpha c \vec{B} \nabla \theta
\\
\ddot {\delta} + \frac{1}{RC} \dot{\delta} + \frac{2eI_c}{\hbar C} \sin \delta &=& \frac{2e}{\hbar C} I \\
P_{a\to \gamma}&=& \frac{1}{16\beta_a}(g_\gamma  \; Bec \;L)^2 \frac{1}{\pi^3f_a^2} \frac{1}{\alpha}
\left( \frac{\sin \frac{qL}{2\hbar}}{\frac{qL}{2\hbar}} \right)^2,
\end{eqnarray}
where
$m_a$ is the axion mass, $f_a$ is the axion coupling constant, $\beta_a=v_a/c$ is the axion velocity,
$\vec{E}$ is the electric field, $\vec{B}$ is the magnetic field,
$g_\gamma =-0.97$ for KSVZ axions, respectively $g_\gamma=0.36$ for DFSZ axions, $q$ is the momentum transfer, $P_{a \to \gamma}$
is the probability of axion decay, $I_c$ is the critical current of junction, and $\alpha\approx 1/137$ is the fine structure constant.

These equations were carefully analysed in \cite{pdu}.
Interestingly, they have a nontrivial solution (besides the trivial solution where
the axion does not interact with the Josephson environment at all): For the
nontrivial solution, inside the weak link area, the Josephson phase angle
$\delta$ synchronizes with the axion misalignment angle $\theta$, self-inducing a formal surface magnetic field
which makes incoming axions decay (but which can also re-create them when leaving the weak link,
i.e. there is a tunneling process). Microscopically, for SNS junctions
this can be interpreted in terms of an axion-- Andreev pair interaction, as sketched
in Fig.~1: Tunneling axions trigger additional Cooper pair transport that would not be there if there were
no axions passing through the junction. Much more details on this
can be found in \cite{prl,pdu}.

\parbox[c]{60mm}{\includegraphics[width=5cm]{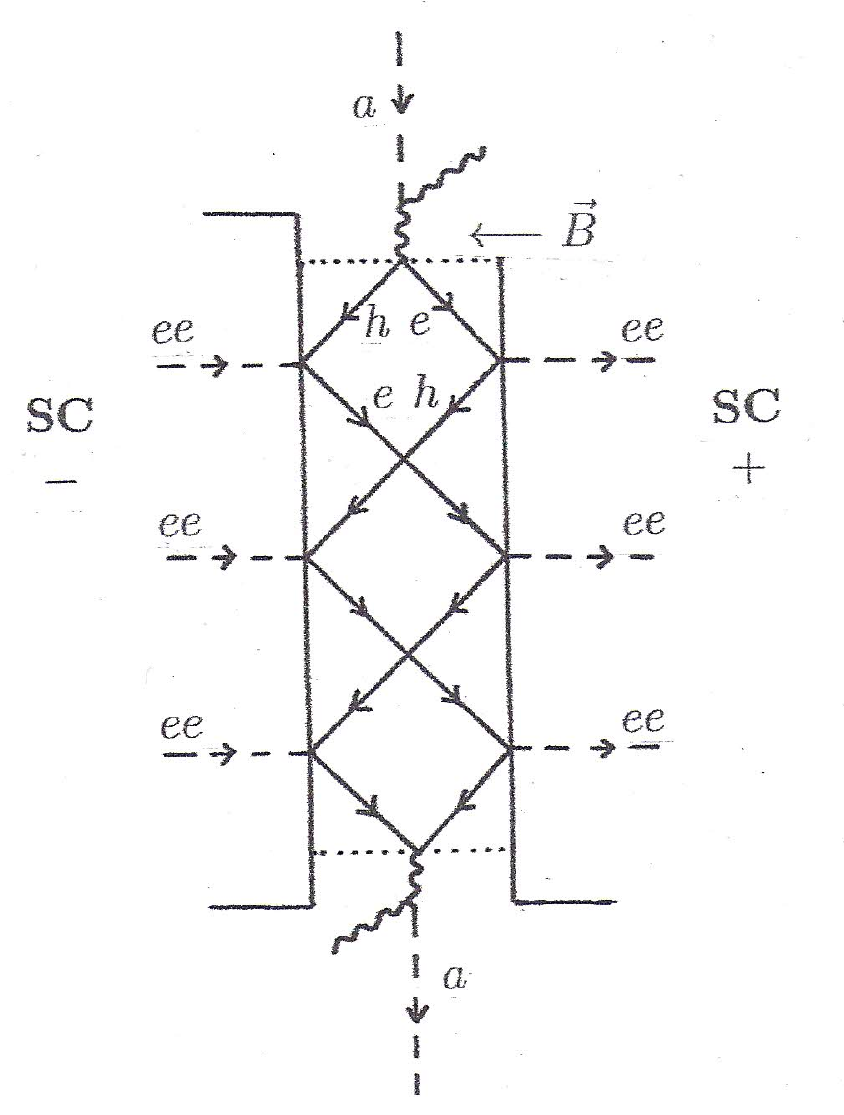}}
\parbox[c]{60mm}{Fig.~1. Microscopic
model of what happens in an SNS junction.
An axion tunnels through the junction and triggers
(by multiple Andreev reflections)
the transport of $n$ Cooper pairs (here $n=3$). Picture from \cite{prl}.}

The result of a longer calculation in \cite{pdu} based on eqs.(2.1)-(2.5)
  is that galactic axion flow through the junction is expected to induce a small oscillating supercurrent in the junction given by
\begin{equation}
I_a(t)=  \sqrt{ \frac{\rho_a v_a}{h \alpha}} w \cdot 2e \cdot cos (\omega_at),
\end{equation}
where $\rho_a$ is the
axion energy density surrounding the Earth, $v_a$ is the axion velocity, $w$ is the width of the Josephson junction, $h=2\pi \hbar$.
Note that the final result (2.6) is independent of $f_a$, $g_\gamma$ and $\Gamma$, which drop out of the equations when the calculations are done.
This axion-induced current produces a Shapiro step (a tiny step (or wiggle)) in the I-V curve that occurs at a voltage $V_a$
where the Josephson frequency $\omega_J$ coincides with the axion mass $m_a$:
\begin{equation}
\hbar \omega_J =2eV_a =m_ac^2=\hbar \omega_a.
\end{equation}

Observing such a wiggle (and excluding any other cause of it) one can thus determine the axion mass.
And, making further assumptions on the number of Andreev reflections involved in Fig.~1, one
can conclude from the wiggle intensity onto the product of axion density $\rho_a$ and
axion velocity $v_a$, though there are large theoretical model uncertainties in this last step,
depending on the microscopic model used.
Overall, the effect of axion flow is mathematically similar to the existence of a second Josephson junction in
addition to the measuring one, with a formal critical current
$I_c^a=  \sqrt{ \frac{\rho_a v_a}{h \alpha}} w \cdot 2e$, interacting in a SQUID-like configuration.
Putting in realistic values for $\rho_a$, $v_a$, and $w$ one notices that the expected
effect from galactic axion flow is a small but perfectly measurable effect: $I_c^a \sim 10^{-8}$A.
In addition, there are also broad-band $1/f$ noise effects from axion flow at very low frequencies, see \cite{fluxnoise} for more details of that aspect.


\section{Candidate axion signals}\label{sec4}

A small measured peak of unknown origin observed at voltage $V_a\approx 55 \mu$V by Hoffmann et al. \cite{ex1}
 was used in \cite{prl} to conclude onto an axion mass of about $m_ac^2 \approx 110 \mu$eV
and an axion density of the order of magnitude $\rho_a \sim 10^{-1} GeV/cm^3$, the latter number being just a rough order-of-magnitude
estimate obtained for an assumed velocity of $2.3 \cdot 10^5$ m/s (the velocity of the Earth relative to the axion background).
This was the very first estimate of this kind. In subsequent work \cite{pdu} it became clear that several other
experiments done with various other types of Josephson junctions also saw unexplained anomalies at a similar voltage.
These Shapiro-step like features were observed at the following voltages (for details, see \cite{pdu}):

\begin{itemize}

\item Hoffmann et al. \cite{ex1} $V_a=(55\pm 1)\mu$V

\item Golikova et al. \cite{ex2}  $V_a=(52\pm 5)\mu$V

\item He et al. \cite{ex3} $V_a=(53 \pm 3)\mu$V

\item Bae et al. \cite{ex4} $V_a=(55 \pm 1)\mu$V

\end{itemize}

To this list one could also add a paper by
Bretheau et al.\cite{ex5} who observe an anomalous current peak
at

\begin{itemize}

\item Bretheau et al. \cite{ex5} $V_a=(50 \pm 3) \mu$V

\end{itemize}

The size of the observed peak ($\sim 10^{-8}$A) is again consistent with
an axion density surrounding the Earth of about $\sim 10^{-1} GeV/cm^3$ (about a quarter of the expected
dark matter density in the halo), again with large theoretical uncertainties.
Of course we should mention at this stage that we cannot
exclude other causes of the observed anomalies in the above-mentioned Josephson experiments, but assuming they are due to axion
 flow then the observed peak positions in the 5 independent experiments point
 quite robustly to an axion mass of
\begin{equation}
 m_ac^2=2eV_a=(106 \pm 6)\mu eV. \label{star}
\end{equation}
The above value is obtained by taking the average over the above 5 condensed matter experiments, with
equal weighting to each experimental group, and using
a conservative estimate of statistical and systematic errors that are expected to occur in these types of experiments.
Our suggestion is that future axion haloscope experiments,
such as ORPHEUS \cite{rybka}, MADMAX \cite{madmax} or ORGAN \cite{organ}, which are
of course extremely important since they are {\em not} based on the
assumption of the validity of the axionic Josephson effect,
should start their search in the mass region (\ref{star}), which is most promising. This suggestion has been taken
up in \cite{organ}.

 \section{A theoretical prediction for a `vanilla' axion}

The above results leading to Eq.~(\ref{star}) are purely experimental. They need to be confirmed in future experiments that are performed
under controlled conditions, i.e. being totally shielded from any potentially disturbing electromagnetic radiation,
and checking for a possible yearly modulation of the signal, as suggested in \cite{prl}.
But can we give a theoretical prediction of the QCD axion mass value? Not really, but at least we
can be inspired by some numerical observation. The QCD axion mass is given by \cite{11}
\begin{equation}
m_ac^2 =57.0(7) \mu eV \cdot  \frac{10^{11}GeV}{f_a} \label{here}
\end{equation}
where $f_a$ is the symmetry breaking scale of Peccei Quinn symmetry. Now the QCD axion really is relevant for {\em all}
types of interactions, it solves the strong CP problem for QCD, it can interact with electromagnetic fields via
the term $\vec{E} \vec{B}$ and thus knows electroweak interactions, and it has gravitational effects as dark matter.
This motivates us to define a 'vanilla' axion (i.e. a most plausible axion) as having a Peccei Quinn symmetry breaking
 scale $f_a$ that is the geometric average of the two most important symmetry breaking scales we know about,
 electroweak symmetry breaking at $v_{EW}=246$ GeV and Planck scale symmetry breaking at $m_{Pl}=1.221 \cdot 10^{19}$ GeV.
 This means the vanilla axion is defined as having the coupling constant
 \begin{equation}
 f_a =\sqrt{v_{EW} m_{Pl}}= 5.48 \cdot 10^{10} GeV \label{there}
 \end{equation}
 (based on a seesaw mechanism of the two most important symmetry breaking scales in nature) and consequently, by combining eq.~(\ref{here}) and (\ref{there}), its mass is theoretically predicted as
 \begin{equation}
 m_ac^2=(104.0\pm 1.3) \mu eV.
\end{equation}
The 5 independent measurements of $m_ac^2=2eV_a$ seen in the Josephson junction experiments above seem to indicate that the
QCD axion could be of vanilla type.

\end{document}